\documentclass[aps,prb,twocolumn,superscriptaddress,showpacs,showkeys,floatfix]{revtex4}

\usepackage{graphicx,color}
\newcommand{\jwj}[1]{\textcolor{blue}{#1}}
\graphicspath{{figs/}}
\begin{document}
\title{Preserving the Q-Factors of ZnO Nanoresonators via Polar Surface Reconstruction}
\author{Jin-Wu Jiang}
    \affiliation{Institute of Structural Mechanics, Bauhaus-University Weimar, Marienstr. 15, D-99423 Weimar, Germany}
\author{Harold S. Park}
    \altaffiliation{Electronic address: parkhs@bu.edu}
    \affiliation{Department of Mechanical Engineering, Boston University, Boston, Massachusetts 02215, USA}
\author{Timon Rabczuk}
    \altaffiliation{Electronic address: timon.rabczuk@uni-weimar.de}
    \affiliation{Institute of Structural Mechanics, Bauhaus-University Weimar, Marienstr. 15, D-99423 Weimar, Germany}
    \affiliation{School of Civil, Environmental and Architectural Engineering, Korea University, Seoul, South Korea }

%\date{22 December 2009}
\date{\today}
\begin{abstract}

We perform molecular dynamics simulations to investigate the effect of polar surfaces on the quality (Q)-factors of zinc oxide (ZnO) nanowire-based nanoresonators.  We find that the Q-factors in ZnO nanoresonators with free polar (0001) surfaces is about one order of magnitude higher than in nanoresonators that have been stabilized with reduced charges on the polar (0001) surfaces.  From normal mode analysis, we show that the higher Q-factor is due to a shell-like reconstruction that occurs for the free polar surfaces.  This shell-like reconstruction suppresses twisting motion in the nanowires such that the mixing of other modes with the resonant mode of oscillation is minimized, and leads to substantially higher Q-factors in the ZnO nanoresonators with free polar surfaces.
\end{abstract}

\maketitle
\pagebreak

\section{introduction}
Nanomechanical resonators (NMR) have, due to their low mass and large surface area for adsorption, received significant attention for their potential as chemical or mass sensors.\cite{EkinciKL,ArletJL,EomK} It has been demonstrated by at least three experimental groups that NMR-based mass sensors can detect mass variations at the level of single atoms.\cite{JensenK,LassagneB,ChiuHY} The sensing performance of NMRs is closely related to the quality (Q)-factor, which reflects the energy dissipated for each vibrational cycle of the nanoresonator. The Q-factor can be affected by external attachment energy loss,\cite{SeoanezC,KimSY2009apl} intrinsic nonlinear scattering mechanisms,\cite{AtalayaJ} or the effective strain mechanism,\cite{JiangJW2012nanotechnology} etc. For graphene with free edges, the Q-factor is significantly reduced by the imaginary localized edge modes.\cite{JiangJW2012jap,KimSY2009nl}

For nanowire (NW)-based NMRs, surface effects due to the large surface to volume ratio have a significant influence on the Q-factor. These surface effects may become even more important in ZnO NWs if the surfaces are polar (0001) surfaces.  Specifically, if ZnO is truncated in the [0001] and [000$\bar{1}$] directions, a Zn ion layer will be exposed at one end while an O ion layer will be exposed at the opposite surface.\cite{WanderA} The two exposed surfaces are strongly polarized, and thus are unstable unless stabilized through one of several possible mechanisms.  In clean environments, the polar surfaces prefer a shell-like reconstruction by outward relaxation of the outermost ion layers.\cite{JedrecyN,KulkarniAJ2005} In another stabilization mechanism, the polar (0001) surfaces can become stable with the saturation of the surface bonds.\cite{WanderA,CarlssonJM,LauritsenJV} A third stabilization mechanism is to transfer some negative charge from the (000$\bar{1}$)-O polar surface to the (0001)-Zn polar surface, which results in the reduction of the charges on the polar surfaces by a factor of 0.25.\cite{NogueraC,DaiS2011jap,DaiS2013jmps}

The polar surface have been found to strongly impact the various physical and chemical properties of ZnO NWs, including the mechanical properties,\cite{KucheyevSO,ChenCQ,ZhangL2006apl,KulkarniAJ,AgrawalR} the electronic or optical band structure,\cite{LinKF,LinKF2006apl,LanyS} and the piezoelectric properties,\cite{MitrushchenkovA,DaiS2013jmps,DaiS2011jap} and others. However, this polar surface effect on the Q-factors of ZnO NW-based NMRs has not been investigated yet, and is thus the focus of the present work.

In this paper, we comparatively study the Q-factors of ZnO NWs under different polar (0001) surface conditions using classical molecular dynamics (MD) simulations: with reduced surface charges on the polar surfaces and with free polar surfaces .  We report significantly higher Q-factors in ZnO NWs with free polar surfaces. Specifically, the Q-factor in ZnO NWs with free polar surfaces is nearly one order of magnitude higher than for ZnO NWs with reduced surface charges.  A normal mode analysis demonstrates that the underlying mechanism is a surface reconstruction to a shell-like structure that occurs on the polar (0001) surfaces.  This shell-like surface reconstruction suppresses twisting motion in the NW, thus minimizing mixing of other modes with the resonant mode of oscillation, and leads to substantially higher Q-factors in the ZnO NMRs with free polar (0001) surfaces.

\section{structure and simulation details}
\begin{figure*}[htpb]
  \begin{center}
    \scalebox{0.7}[0.7]{\includegraphics[width=\textwidth]{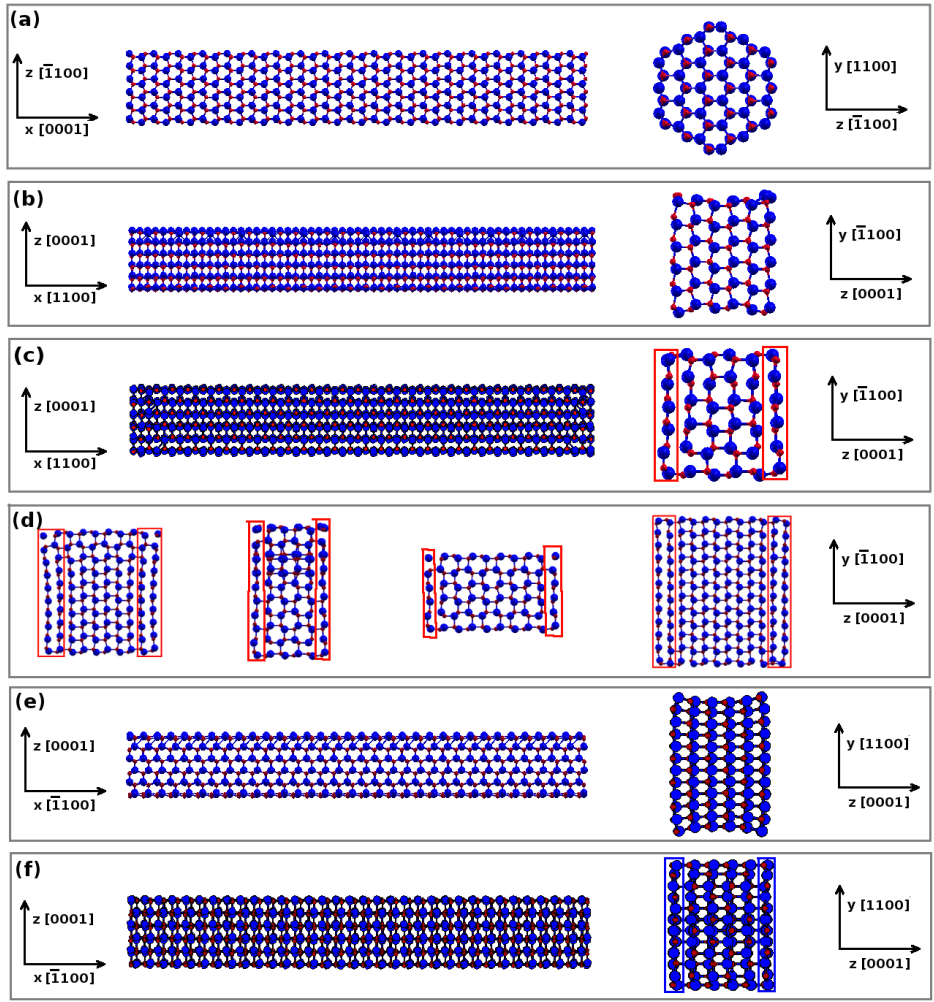}}
  \end{center}
  \caption{(Color online) The optimized configuration of ZnO NWs studied in the present work. (a) Hexagon NW: [0001]-oriented with hexagonal cross section. There are no polar surfaces in this NW. Side view is on the left and the cross section is on the right. (b) Reduce NW: [1100]-oriented. The charges for ions on the polar surfaces are reduced by a factor of 0.25.  (c) Free NW: [1100]-oriented with free polar surfaces in the $z$ direction. The free polar surfaces are reconstructed into shell-like structure (highlighted by red box). (d) The relaxed cross section for [1100]-oriented free NW of various sizes, all showing the shell-like structure. (e) Reduce NW: [1$\bar{1}$00]-oriented. The charges for ions on the polar surfaces are reduced by a factor of 0.25\cite{NogueraC,DaiS2011jap,DaiS2013jmps}. (f) Free NW: [1$\bar{1}$00]-oriented with free polar surfaces in the $z$ direction. The polar surface (highlighted by blue box) is not a distinct shell-like structure.}
  \label{fig_cfg}
\end{figure*}

We studied the Q-factors of ZnO NMRs via classical MD simulations. The interatomic interactions were described by the following Buckingham-type potential:
\begin{equation}
V_{\rm tot}(r_{ij}) = \sum_{i\not=j}^{N} A \exp(-\frac{r_{ij}}{\rho}) - \frac{C}{r^{6}_{ij}} + V_{\rm long}(r_{ij}),
\label{eq_buckingham}
\end{equation}
where $N$ is the total number of ions. The short-range parameters $A$, $C$, and $\rho$ for O were developed by Catlow,\cite{CatlowCRA} while the other short-range parameters are from Catlow and Lewis.\cite{LewisGV} \jwj{$V_{\rm long}(r_{\rm ij})=q_{i} q_{j}/r_{ij}$ is the long-ranged Coulombic interaction, where $q_{i}$ and $q_{j}$ are the atomic charges.}  All ZnO NWs in our study have a hexagonal wurtzite crystal structure, where for bulk ZnO the lattice constant of the hexagonal lattice is $a=$3.2709~{\AA}, while the lattice constant in the perpendicular direction of the hexagonal plane is $c=$5.1386~{\AA}. The intra-cell geometrical parameter is $u=0.3882$.

Several efficient techniques have been developed to deal with the summation for the long-range electrostatic interaction between ions, i.e. $V_{\rm long}$ in Eq. \ref{eq_buckingham}. The Ewald summation is the traditional way to calculate the electrostatic interactions within a bulk system with periodic boundary conditions.\cite{EwaldPP} However, for surface-dominated nanomaterials such as the NWs in the present work, the periodicity assumption is clearly violated, and therefore it is crucial to employ a truncation-based summation method.  In this work, we have utilized the truncation-based summation approach initially proposed by Wolf et al. in 1999\cite{WolfD} and further developed by Fennell and Gezelter in 2006.\cite{FennellCJ}  A key development in the work by~\citet{FennellCJ} was to ensure that the electrostatic force and potential are consistent with each other, while remaining continuous at the interatomic cut-off distance.  In a recent study, Gdoutos et al. have quantified the errors in using the traditional periodic Ewald summation approach for ZnO NWs.\cite{GdoutosEE} In our calculation, we have chosen the damping parameter $\alpha=0.3$~{\AA$^{-1}$} and the cut-off $r_{c}=10.0$~{\AA}, which gives convergent results for both the piezoelectric\cite{DaiS2011jap} as well as the mechanical properties of ZnO NWs.\cite{AgrawalR}

The standard Newton equations of motion are integrated in time using the velocity Verlet algorithm with a time step of 1 fs. Both ends of the NW are held fixed to simulate fixed/fixed boundary conditions, while free boundary conditions are applied in the two lateral directions.  Our simulations are performed as follows.  First, the Nos\'e-Hoover\cite{Nose,Hoover} thermostat is applied to thermalize the system to a constant temperature within the NVT (i.e. the number of particles N, the volume V and the temperature T of the system are constant) ensemble, which is run for 50~{ps}.  Free mechanical oscillations of the ZnO NMR is then actuated by adding a velocity distribution to the system, which follows the morphology of the first bending mode of the ZnO NW~\cite{JiangJW2012jap}. The imposed velocity distribution, or actuation energy, is $\Delta E=\alpha E_{k}^{0}$, where $E_{k}^{0}$ is the total kinetic energy in the NMR after thermalization but just before its actuation and $\alpha=0.5$ is the actuation energy parameter.  After the actuation energy is applied, the system is allowed to oscillate freely within the NVE (i.e. the particles number N, the volume V and the energy E of the system are constant) ensemble for more than 1000~{ps}. The data from the NVE ensemble is used to analyze the mechanical oscillation, and the Q-factors of the ZnO NMR. \jwj{The decay of the oscillation amplitude for the kinetic energy is used to extract the Q-factor by fitting the kinetic energy from the NVE ensemble to a function $E_{k}(t)=a+b(1-2\pi/Q)^{t} \cos(\omega t)$, where $\omega$ is the frequency, $a$ and $b$ are two fitting parameters and $Q$ is the resulting quality factor. }

Fig.~\ref{fig_cfg} shows the relaxed, energy-minimized configuration of the ZnO NWs we considered in this work, where the energy minimized configuration is obtained using the conjugate gradient method. All NWs have similar dimensions, so that a consistent comparison between different orientations and surface treatments can be performed. The NW in panel (a) is [0001]-oriented with hexagonal cross section. The dimension is $97.6\times 19.6\times 17.0$~{\AA}. The $y$ axis is along the $[1100]$ direction, while the $z$ axis is along the $[\bar{1}100]$ direction. There are no polar surfaces in this NW. We will refer to these NWs as hexagon ZnO NWs in the following.

\begin{figure}[htpb]
  \begin{center}
    \scalebox{1.0}[1.0]{\includegraphics[width=8cm]{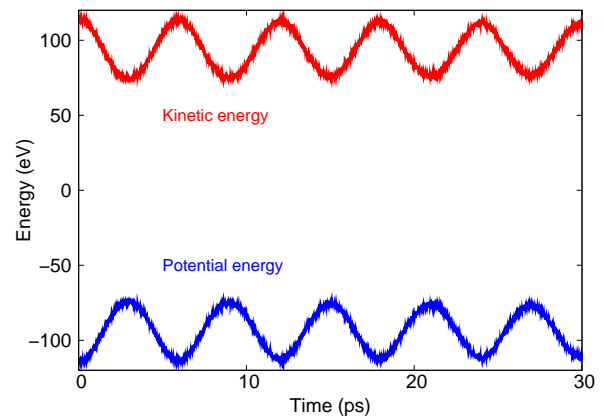}}
  \end{center}
  \caption{(Color online) The energy exchange between the kinetic and potential energy during the resonant oscillation at 300~K for [1100]-oriented ZnO NWs of dimension $98.1\times 17.0 \times 15.4$~{\AA} with reduced charges on the polar (0001) surfaces. The potential energy has been shifted so that the total energy is zero. }
  \label{fig_energy_conserve}
\end{figure}

\begin{figure}[htpb]
  \begin{center}
    \scalebox{0.8}[0.8]{\includegraphics[width=8cm]{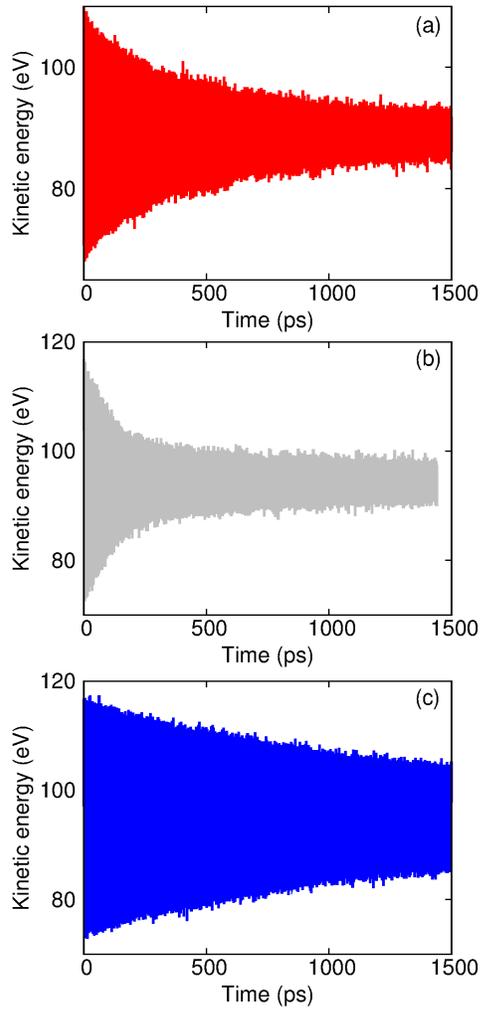}}
  \end{center}
  \caption{(Color online) The kinetic energy time history at 300~K for (a) [0001]-oriented hexagon ZnO NW, (b) [1100]-oriented reduce NW, and (c) [1100]-oriented free NW.}
  \label{fig_kinetic_energy}
\end{figure}

\begin{figure}[htpb]
  \begin{center}
    \scalebox{1.0}[1.0]{\includegraphics[width=8cm]{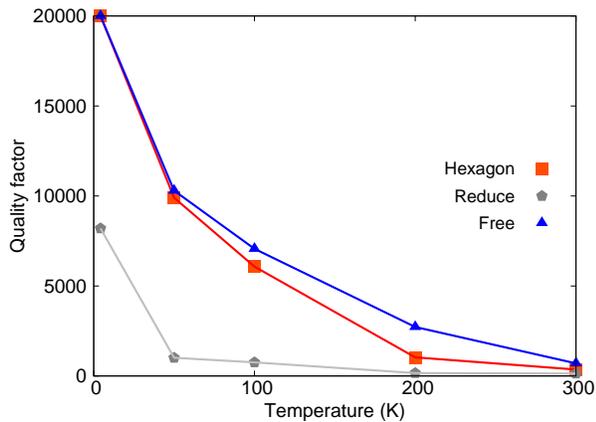}}
  \end{center}
  \caption{(Color online) The temperature dependence for the quality factor of the [0001]-oriented hexagon NW, [1100]-oriented reduce NW, and [1100]-oriented free ZnO NW.}
  \label{fig_qfactor}
\end{figure}
\begin{figure}[htpb]
  \begin{center}
    \scalebox{1.1}[1.1]{\includegraphics[width=8cm]{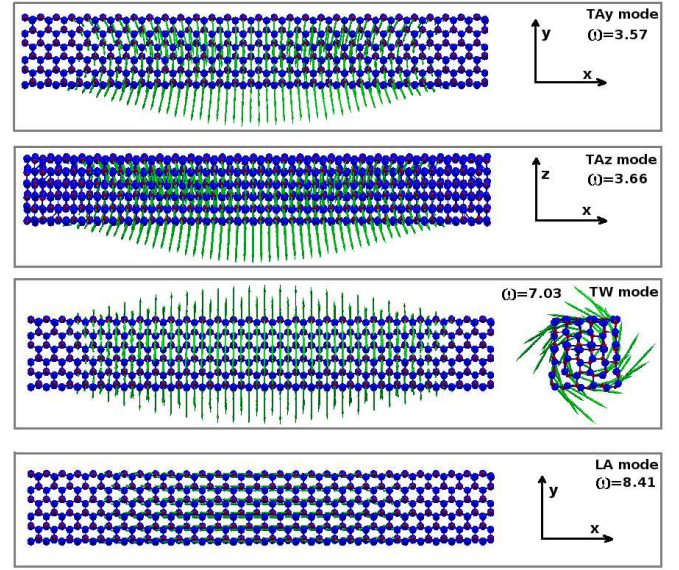}}
  \end{center}
  \caption{(Color online) The polarization vector of the four acoustic phonon modes in the free ZnO NW. From top to bottom: TA$_{y}$, TA$_{z}$, TW, and LA modes. Arrows in the figure represent the polarization vector of the phonon mode. Numbers in the figure are the frequency for each mode in the unit of cm$^{-1}$.}
  \label{fig_u}
\end{figure}

\begin{figure*}[htpb]
  \begin{center}
    \scalebox{0.76}[0.76]{\includegraphics[width=\textwidth]{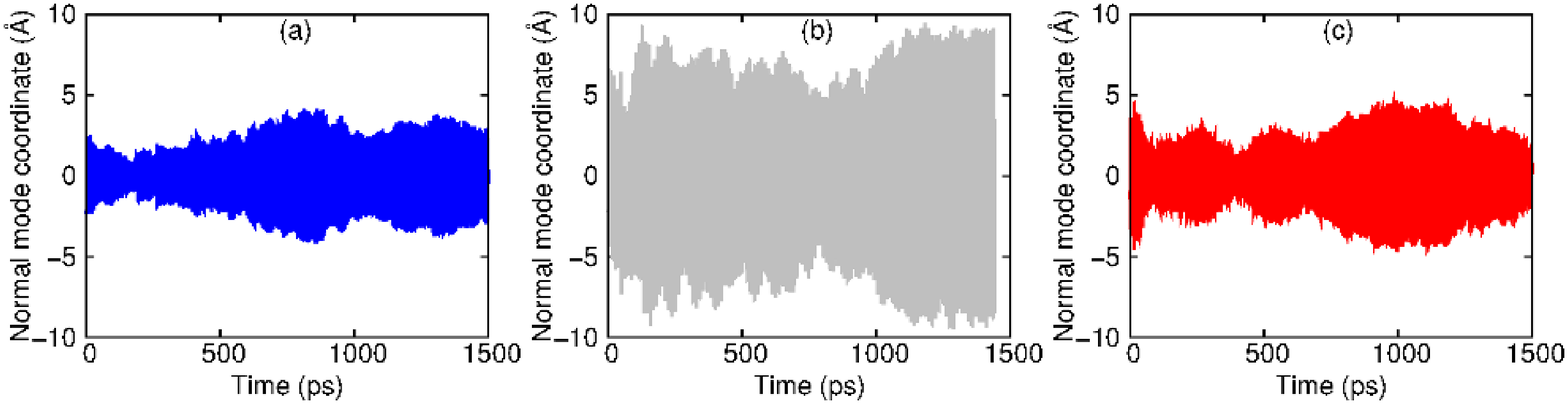}}
  \end{center}
  \caption{(Color online) The normal mode coordinates at 300~K for the TW modes in (a) [0001]-oriented hexagon ZNO NW, (b) [1100]-oriented reduce NW, and (c) [1100]-oriented free NW.}
  \label{fig_Qt_TW}
\end{figure*}

The NW in panel (b) is [1100]-oriented with dimension $98.1\times 17.0 \times 15.4$~{\AA}. The $y$ axis is along the $[\bar{1}100]$ direction, while the $z$ axis is along the [0001] polar direction. The side view is shown on the left while the cross section is on the right. The surface charges on the two polar surfaces are reduced by multiplying by a  factor of 0.75. More specifically, the charge of the Zn ions on the (0001)-Zn surface are reduced from $+2e$ to $+1.5e$, while the charge of the O ions on the $(000\bar{1})$-O surface are reduced from $-2e$ to $-1.5e$.  This charge reduction is justified by the fact that surface Zn and O atoms have only three nearest neighbors as compared to bulk Zn and O, which have four.  Hence, a 25\% reduction in the surface charges can be utilized to stabilize the polar (0001) surfaces.\cite{NogueraC} \jwj{A more rigorous physical and mathematical justification can be found in~\citet{NogueraC}.  In that work, from electrostatic and electronic structure calculations, it was found that a counterfield can be created to quench the macroscopic dipole moment if the Zn-terminated polar surface is less positive and the O-terminated polar surface is less negative by a factor of $R_{2}/(R_{1}+R_{2})\approx 0.75$, where $R_{1}=0.69$~{\AA} and $R_{2}=1.99$~{\AA} are the alternating distance for ion layers in the (0001) lattice direction.}  It is clear from panel (b) that the crystal structure of the two polar surfaces is maintained after the energy minimization. In the following, we will refer to these NWs as reduce ZnO NWs.

The ZnO NW in panel (c) has the same orientation and cross sectional dimensions as the free ZnO NW in (a). The only difference is that the surface charges on the two polar surfaces are kept unchanged. The cross sectional view clearly shows a shell-like reconstruction at the two polar surfaces (highlighted by red boxes).  The Zn and O ions move into an essentially planar configuration which suppresses the surface polarization.\cite{DaiS2013jmps}  In this way, the outermost planes of atoms in the $z$ direction try to peel away from the bulk system, leading to shell-like layers on the (0001) and $(000\bar{1})$ surfaces. The shell-like reconstruction induces the surface to move away from the center of the NW, and thus the bonding of the shell-reconstructed (0001) surfaces to the bulk occurs weakly through a reduced number of Zn-O bonds. The shell-like reconstruction also happens in ZnO NWs of various larger cross sections as shown in panel (d). We will refer to the NW shown in panel (c) for the remainder of this work as the free ZnO NW.

The NW in panel (e) is [1$\bar{1}$00]-oriented ZnO NWs with dimension $96.3\times 19.6 \times 15.4$~{\AA}. The $y$ axis is along the $[1100]$ direction, while the $z$ axis is along the [0001] polar direction. The surface charges on the two polar surfaces are reduced by multiplying by a factor of 0.75. The ZnO NW in panel (f) has the same orientation and cross sectional dimensions as the free ZnO NW in (e). The only difference is that the surface charges on the two polar surfaces are kept unchanged. There is no obvious shell-like surface in this free NW. 

\jwj{It is important to note that the ZnO NW configurations we have chosen to study were done so because they are the most commonly synthesized and studied configurations experimentally.  First of all, the hexagon NW are commonly synthesized experimentally, particularly when ZnO NWs are grown vertically from a substrate, like sapphire, which possess six or three-fold symmetry.  Thus, the resulting epitaxially grown ZnO NWs prefer to have a hexagonal cross section.\cite{YangP2002} The free NWs are typically grown by sublimation of ZnO powder without introducing a catalyst.\cite{WangZL} An interesting phenomenon in the free NW is experimentally-observed planar defects parallel to the (0001) polar surfaces.\cite{DingY} The shell-like surface reconstruction in Fig.~\ref{fig_cfg}~(a) is also a planar defect and shares some similarities with this experimentally-observed planar defect. For the reduce NW, the charge reduction on the (0001) polar surface is one of the most promising mechanisms to stabilize the polar surface. This stabilization mechanism was proposed based on both electrostatic considerations\cite{NoskerRW} as well as electronic structure calculations\cite{ZwanzigR}.  Many experimental efforts have been devoted to investigate the actual process for this stabilization mechanism. For example, the charge reduction can be realized via the transfer of charges between (0001)-Zn and (000$\bar{1}$)-O polar surfaces (i.e., electronic relaxation).\cite{DulubO,LauritsenJV} }

\section{results and discussion}

\jwj{Fig.~\ref{fig_energy_conserve} shows the energy exchange between the potential and kinetic energy during the resonant oscillation in the NVE ensemble at 300~K for the [1100]-oriented reduce ZnO NW. The potential energy has been overall shifted, so that the total energy is zero.} 

Fig.~\ref{fig_kinetic_energy}~(a) compares the kinetic energy time history of the hexagon, reduce, and free ZnO NWs at 300~K. The oscillation in the kinetic energy reflects the resonant oscillation in the ZnO NWs. The decay of the oscillation amplitude in the kinetic energy represents the energy dissipation in the NMR. Panel (a) shows the kinetic energy for the hexagon NW. \jwj{ The hexagon NW shows a higher Q-factor, i.e. lower energy dissipation, than the reduce ZnO NWs.  The reason for this is the fact that the hexagon NW does not have any polar surfaces.  Because of this, its surfaces are relatively smooth and stable, and do not undergo complex relaxations like those seen in the reduce NWs.  The more orderly surface structure is important because it enables the surface to oscillate more coherently with the bulk, which leads to less energy dissipation and a higher Q-factor.\cite{KimSY2008prl,JiangJW2012jap}).  We would also naively expect that the hexagon NW would also have a higher Q-factor than the free NW.  However, interestingly enough, panel (c) shows that the rate of energy dissipation in the free NW is quite comparable with and actually exceeds that in the hexagon NW. This result indicates that the polar surface does not necessarily lead to low Q-factor. }

Fig.~\ref{fig_qfactor} shows the temperature dependence of the Q-factor in hexagon, reduce, and free ZnO NWs. The Q-factor of the free NW is very close to the reduce NW. More precisely, it is higher than the Q-factor of the reduce NW. The Q-factor of the free NW is about one order of magnitude higher than the Q-factor of the reduce NW across the whole temperature range.

To understand the remarkable preservation effect that polar surfaces have on the Q-factors of the free ZnO NWs, we examine the quite different surface structure of this NWs as compared with the reduce NW. A novel feature in the free NW is the shell-like surfaces resulting from the reconstruction of the polar (0001) surfaces, whereas there is no shell-like reconstruction in the reduce NWs.  Because of this, our key insight is that the shell-like surface structure may act as a safeguard for the resonant oscillation in the free ZnO NW. More specifically, we will demonstrate that the shell-like surfaces prevent the ZnO NW from exhibiting torsional motion during the resonant mechanical oscillation. \jwj{ There are four types of low-frequency acoustic phonon modes in the quasi-one-dimensional ZnO NW: two bending modes (TA$_{y}$ and TA$_{z}$), one twisting (TW) mode, and one longitudinal (LA) mode. The frequency of these four phonon modes are in the order: TA$_{y}$$<$TA$_{z}$$<$TW$<$LA mode. Fig.~\ref{fig_u} shows the frequency and vibration morphology (i.e eigen vector) of the lowest-frequency modes from these four different types of phonon modes in the free ZnO NW.  No imaginary frequencies are observed, which illustrates the stability of relaxed structures in our study. Among these modes, only the torsional mode is able to strongly interact with the bending-like resonator oscillation due to its special vibrating morphology, which satisfies the symmetry selection rules in the phonon-phonon scattering process.\cite{BornM}} As a result, energy dissipation in the free ZnO NW is much lower, which directly results in higher Q-factor.

\begin{figure}[htpb]
  \begin{center}
    \scalebox{1.0}[1.0]{\includegraphics[width=8cm]{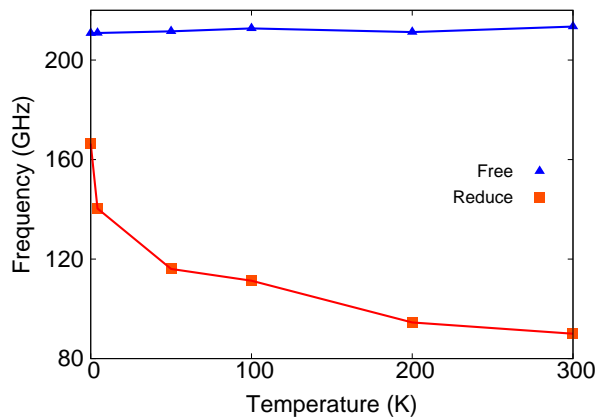}}
  \end{center}
  \caption{(Color online) The frequency of the TW modes in free and reduce NWs. The frequency at 0~K is calculated from the eigenvalue solution of the dynamical matrix. The frequencies at finite temperature are obtained from the time history of the normal mode coordinate for the TW modes.}
  \label{fig_tw_mode}
\end{figure}
\begin{figure}[htpb]
  \begin{center}
    \scalebox{1.0}[1.0]{\includegraphics[width=8cm]{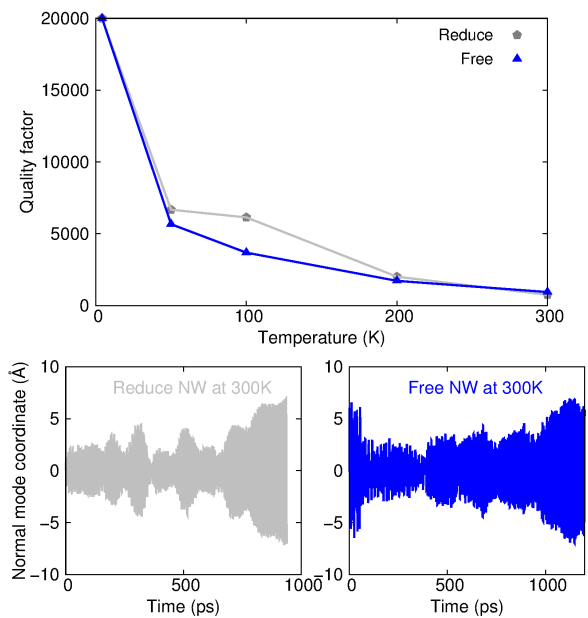}}
  \end{center}
  \caption{(Color online) The resonant behavior of the [1$\bar{1}$00]-oriented ZnO NWs. Top panel: The quality factors in free and reduce NWs are close to each other. Bottom panels: The normal mode coordinates at 300~K for the TW modes in the reduce NW (left, gray online) and the free NW (right, blue online).}
  \label{fig_qfactor_1m100}
\end{figure}

To verify the key role of the shell-like surface reconstruction in preserving the Q-factor, we calculate the normal mode coordinate for the twisting (TW) mode,
\begin{equation}
Q^{\rm TW}(t)=\sum_{j=1}^{3N}\xi^{\rm TW}_{j}(r_{j}(t)-r_{j}^{0}),
\label{eq_Qt}
\end{equation}
where $k$ is the mode index, $3N$ is the total degrees of freedom in the NW, $r_{j}(t)$ is the trajectory of the $j$-th degree of freedom from the MD simulation, $r_{j}^{0}$ is the corresponding equilibrium position and {\boldmath $\xi$}$^{\rm TW}=(\xi^{\rm TW}_{1}, \xi^{\rm TW}_{2}, \xi^{\rm TW}_{3}, ..., \xi^{\rm TW}_{3N})$ is the polarization vector of the TW mode. The polarization vectors can be obtained by solving the eigenvalue problem of the dynamical matrix, which is derived from the interatomic potential. The dynamical matrix is obtained by $K_{ij}=\partial^{2}V/\partial x_{i}\partial x_{j}$, where $V$ is the interatomic potential and $x_{i}$ is the position of the i-th degree of freedom. This formula is realized numerically by calculating the energy change after a small displacement of the i-th and j-th degrees of freedom. Both ends of the NW are fixed in this eigenvalue calculation, which is consistent with the fixed boundary condition applied in the above simulation of the mechanical resonance.  The top panel in Fig.~\ref{fig_u}~(c) displays the polarization vector ({\boldmath $\xi$}$^{\rm TW}$) of the TW mode in the free ZnO NW. \jwj{It should be noted that the eigenvalue problem has been solved prior to the MD simulation. Using the eigenvector of the normal mode, the normal mode coordinate for the twisting mode is evaluated according to its definition in Eq.~(\ref{eq_Qt}) at each MD simulation step for the resonator oscillation. The periodic boundary condition is applied during the structural optimization of the ZnO nanowire. After this optimization, both ends are fixed both in the simulation of the resonator oscillation and in the calculation of the phonon modes. The fixed boundary condition in the axial direction is used in our simulation so as to mimic the actual boundary condition in the experiment, where the nanoresonators are often clamped at both ends.\cite{BunchJS2007sci,LassagneB}}

The normal mode coordinate for the TW mode is useful as it elucidates the relative importance of the torsional mode to the overall motion of the NW. If the amplitude of the normal mode coordinate for the TW mode is large, it means that the corresponding torsional motion is energetically favorable for the NW. The time history of the normal mode coordinate at room temperature for the TW modes is shown in the three panels in Fig.~\ref{fig_Qt_TW} for the hexagon ZnO NW (left), the reduce ZnO NW (middle), and the free ZnO NW (right). Indeed, we find that the $Q(t)$ for the TW mode in the free ZnO NW is much smaller than in the reduce ZnO NW. This result means that the twisting movement is weaker in the free ZnO NW, which provides a direct evidence for our hypothesis that the shell-like surfaces in the free NW are indeed able to prevent the NW from exhibiting large torsional motion.

The frequency of the TW mode provides another direct proof for the above torsion-related preserving mechanism. Fig.~\ref{fig_tw_mode} shows the frequency of the TW modes in the free and reduce ZnO NWs. The frequency at 0~K is obtained from the solution of the eigenvalue problem for the dynamical matrix. The frequency value at finite temperature is extracted from the oscillation frequency of the normal mode coordinate for the TW modes as shown in Fig.~\ref{fig_Qt_TW}. \jwj{At finite temperature, different phonon modes interact with each other due to the nonlinear phonon-phonon scattering, which influences the frequency and the life time of each phonon mode.\cite{BoniniN} This is the origin for the temperature-dependence of the mode frequency in the figure.} It shows that the frequencies of the TW modes in the free NW are more than 25\% higher than the reduce NW near 0 K, and the difference is larger at higher temperature. A higher frequency indicates that more energy is required to excite the torsional motion in the free NW; i.e the torsional motion is not energetically favorable for the free NW. In other words, the shell-like polar surface reconstruction prevents the free NW from exhibiting torsional motion by increasing the frequency of the TW mode in the system. \jwj{Furthermore, the frequency of the TW mode in free NW is almost independent of the temperature, which reflects the weak interaction between this mode with other modes (including the resonator bending mode) in the system. Hence, this mode is more difficult to be excited by the resonator oscillation (bending), i.e the torsional motion is prevented.}

In the above, we have established that the shell-like polar (0001) surface reconstruction serves to enhance the Q-factor in the [1100]-oriented ZnO NW with free polar surfaces, because the reconstructed surfaces act like a safeguard for the mechanical resonant oscillation.  However, a natural extension to this is to investigate what the difference in the Q-factors of free and reduce NWs is if the polar (0001) surface reconstructs, but not into a distinctive shell-like structure?  From the self-consistent extension of the above preserving mechanism, the difference in the Q-factor should be small in this situation, because of the absence of the shell-like surfaces preventing the NW from undergoing torsional motion.

To validate this hypothesis, we studied the resonant oscillation in [1$\bar{1}$00]-orientated ZnO NWs with reduce and free polar surfaces. For the free NW, Fig.~\ref{fig_cfg}~(f) shows that the polar surfaces are reconstructed, though not into a shell-like surface structure, as the reconstructed polar surfaces are still strongly connected to the bulk through many Zn-O bonds. This is different from the distinct shell-like surfaces shown in Fig.~\ref{fig_cfg}~(c), where the two reconstructed polar surfaces are only weakly connected to the center part through few Zn-O bonds. As a result, Fig.~\ref{fig_qfactor_1m100} shows that the Q-factors in these two NWs are close to each other, because there is no obvious preserving mechanism in the [1$\bar{1}$00]-orientated free ZnO NW.  Furthermore, the two bottom panels in Fig.~\ref{fig_qfactor_1m100} show that the normal mode coordinates for the TW modes in the reduce and free NWs at 300~K have similar amplitude; i.e the torsional motions in the reduce and free [1$\bar{1}$00]-orientated ZnO NWs are on the same strength level. As a result, the Q-factors in the reduce and free [1100]-oriented ZnO NMRs are quite similar.

Before concluding, we would like to point out that we have focused on NWs with small cross sections, because calculation of the electrostatic term of the interatomic potential is fairly expensive computationally. However, the mechanism that preserves the high Q-factor should also be found in larger cross section NWs, since the shell-like surface reconstruction also happens in thicker NWs as shown in Fig.~\ref{fig_cfg}~(d). \jwj{Furthermore, we find that the formation energy, $\epsilon = \frac{E}{N} - \epsilon_{0}$, for the hexagon, reduce [1100], free [1100], reduce [1$\bar{1}$00], and free [1$\bar{1}$00] nanowires are: 0.18, 1.17, 0.18, 1.16, and 0.16~{eV}, respectively. Here $E$ is the total energy in the nanowire. $N$ is the total atom number in the nanowire. $\epsilon_{0}$ is the energy per atom in the bulk ZnO. According to this comparison, the hexagon and free NWs are most likely to be realized in experiments, which agrees with the experimental findings.\cite{YangP2002,WangZL}  Therefore, it should be possible to observe the preserving mechanism for the high Q-factor in experiments.}

\section{conclusion}

In conclusion, we have performed classical MD simulations to comparatively study the mechanical resonant oscillation in ZnO NWs having two different surface conditions: those with reduced surface charges to stabilize the polar (0001) surfaces and those with free polar (0001) surfaces.  We find that the Q-factor in the free ZnO NW is much higher than the quality factor in the reduce NWs. By comparing the normal mode coordinates of the TW modes, we find that the free polar surfaces undergo a shell-like reconstruction, which prevents the resonant oscillation in the free NW from being dissipated through torsional motion.

\textbf{Acknowledgements} The work is supported by the Grant Research Foundation (DFG) Germany.  HSP acknowledges support from the Mechanical Engineering Department of Boston University.

%\bibliographystyle{aipnum4-1}
%\bibliography{biball}

\end{document}